\documentclass{article}
\usepackage{authblk}
\usepackage[utf8]{inputenc}
\usepackage{graphicx}
\usepackage{amsmath}
\usepackage{diagbox}
\usepackage{amsfonts}

\title{ELPINN: Eulerian Lagrangian Physics-Informed Neural Network}
\date{}
\author[1]{Sukirt Thakur}

\affil[1]{\small{School of Mechanical Engineering,
            Purdue University, 
            West Lafayette,
            47907, 
            Indiana,
            USA}}

\author[2]{Maziar Raissi}

\affil[2]{ Department of  Mathematics,
            University of California, 
            Riverside,
            92521, 
            California,
            USA}

\begin{document}

\maketitle

\begin{abstract}
Physics-Informed Neural Networks (PINNs) have gained widespread popularity for solving inverse and forward problems across a range of scientific and engineering domains. However, most existing PINN frameworks are limited to the Eulerian domain, where physical quantities are described at fixed spatial locations. In this work, we propose a novel PINN-based framework that couples Eulerian and Lagrangian perspectives by using particle trajectory data to reconstruct Eulerian velocity and pressure fields.

We evaluate the performance of our method across three distinct fluid flow scenarios: two-dimensional external flow past a cylinder, two-dimensional internal flow in a confined geometry, and three-dimensional internal flow inside an airplane cabin. In all three cases, we successfully reconstruct the velocity field from Lagrangian particle data. Moreover, for the 2D external and internal flows, we recover the pressure field solely through the physics-informed learning process, without using any direct pressure measurements.

We also conduct a sensitivity analysis to understand the effects of temporal resolution and particle count on the reconstruction accuracy. Our results show that smaller time-step sizes significantly improve the predictions, while the total number of particles has a comparatively smaller influence.

These findings establish the potential of our coupled Eulerian-Lagrangian PINN framework as a powerful tool for enhancing experimental methods such as Particle Tracking Velocimetry (PTV). Looking ahead, this approach may be extended to infer hidden quantities such as pressure in three-dimensional flows or material properties like viscosity, opening new avenues for data-driven fluid dynamics in complex geometries.
\end{abstract}

\section{Introduction}

The application of machine learning (ML) and deep learning (DL) techniques to scientific problems is rapidly expanding. These approaches have been successfully applied across diverse fields, including fluid mechanics \cite{Brunton2021, Duraisamy2019} and epidemiology \cite{Bi2019}. However, many physical and biological systems suffer from sparse, noisy, and difficult-to-acquire data, making traditional ML frameworks prone to overfitting and generalization errors \cite{Karniadakis2021}. 

Physics-Informed Neural Networks (PINNs) \cite{Raissi2019} have emerged as a robust computational framework for solving forward and inverse problems across multiple domains. PINNs effectively handle small datasets by incorporating governing physical equations as a regularization term, reducing the dependency on large labeled datasets. This approach has been successfully applied to fluid dynamics \cite{Cai2021, Raissi2020, Sukirt_PoF_TC},  rheology \cite{viscoelasticNet, Sukirt_fractional_arxiv} structural mechanics \cite{Haghighat2021, Henkes2022}, and heat transfer \cite{Cai2021HT, Lu2021}. Moreover, PINNs have been extended to tackle high-dimensional PDEs \cite{Han2018} and multi-physics problems \cite{Ma2022}. 

Most PINN applications focus on governing equations formulated in the Eulerian domain. In this framework, physical quantities such as velocity, pressure, and temperature are described at fixed spatial locations. This perspective is well-suited for continuum mechanics, where fluid flow and heat transfer are analyzed across a spatial domain. The Eulerian formulation, commonly used to describe systems governed by partial differential equations (PDEs), is particularly effective for fluid dynamics applications, such as solving the Navier-Stokes equations \cite{Karniadakis2021, Sun2020}. 

Conversely, the Lagrangian perspective focuses on individual particles or fluid elements, tracking their trajectories over time \cite{Zefran2003, Maire2007}. This approach is fundamental for studying particle-laden flows \cite{Jesse2013}, turbulence modeling \cite{Charles2011}, molecular dynamics \cite{Khalili2005}, and transport phenomena such as sediment transport \cite{Chen2015}. While PINNs have been widely applied to Eulerian problems, their extension to Lagrangian descriptions remains underexplored. However, many physical systems exhibit hybrid behavior, requiring a coupled Eulerian-Lagrangian approach. 

For instance, sediment transport in a river involves suspended particles (Lagrangian) influenced by fluid flow (Eulerian) and vice versa \cite{Ballio2018}. Similarly, in biological systems, cells within tissues experience fluid forces (Eulerian) while simultaneously migrating and interacting with neighboring cells (Lagrangian). Accurately modeling these interactions necessitates a unified framework that bridges both viewpoints. 

In this work, we present a PINN framework that leverages Lagrangian particle data to make predictions in the Eulerian domain. Specifically, we track passive particles in the Lagrangian domain to reconstruct the velocity field in the Eulerian domain. Using this velocity field, we further infer the pressure distribution by minimizing the residuals of the Navier–Stokes equations. The governing equations and the PINN architecture for Eulerian–Lagrangian coupling are described in Section~\ref{problem_def}.

We evaluate our framework on three test cases of increasing complexity: a two-dimensional external flow past a cylinder, a two-dimensional internal flow in a confined domain, and a three-dimensional internal flow representing airflow within an airplane cabin. In all three cases, we successfully reconstruct the velocity field from particle trajectory data. For the 2D external and internal flow problems, we additionally recover the pressure field using physics-informed learning. The corresponding results and a detailed sensitivity analysis are presented in sections ~\ref{results_fpc} and ~\ref{results_2D} for the 2D cases and in Section~\ref{results_airplane} for the 3D airplane geometry. We conclude with a summary and discussion of the broader implications of our findings in Section~\ref{conclusion}.

\section{Problem definition}
\label{problem_def}
\subsection{Governing equations}

Our objective is twofold: first, to track inert particles and obtain the velocity field using the trajectory of the particles; and second, to use this velocity field in a Lagrangian frame of reference to calculate the pressure field by minimizing the residual of the Navier-Stokes equations. We consider the passive transport of mass-less particles, whose dynamics are governed by the following equation:

\begin{equation}\label{eqn_1}
    \boldsymbol{x}_t = \boldsymbol{u}(t,\boldsymbol{x}),
\end{equation}
where the subscript $t$ denotes the time derivative and $\boldsymbol{x}$ is the spatial coordinate that is $\boldsymbol{x} = (x, y)$ in two dimensions and $\boldsymbol{x} = (x, y, z)$ in three dimensions. The velocity field is defined as $\boldsymbol{u} = (u, v)$ in two dimensions and as $\boldsymbol{u} = (u, v, w)$ in three dimensions. Equation \eqref{eqn_1} can be written in terms of the individual components as:

\begin{equation}
    \frac{dx}{dt} = u,
    \frac{dy}{dt} = v,
    \frac{dz}{dt} = w.
\end{equation}

The conservation of mass for an incompressible fluid is given by:

\begin{equation}\label{continuity}
    \nabla \cdot \boldsymbol{u} = 0,
\end{equation}
where $\boldsymbol{u}$ is the fluid velocity vector. In terms of individual velocity components, the conservation of mass in two-dimensions can be written as:

\begin{equation}
    u_x + v_y = 0.    
\end{equation}

The conservation of momentum for an incompressible Newtonian fluid under isothermal, single-phase, transient conditions in the absence of a body force is given by:

\begin{equation}\label{momentum}
    \left( \frac{\partial \boldsymbol{u}}{\partial t} + \boldsymbol{u}\cdot \nabla \boldsymbol{u}\right) = -\frac{1}{\rho}\nabla p + \nu  \nabla^2 \boldsymbol{u},
\end{equation}
where $\rho$ is the density of the fluid, $\boldsymbol{u}$ is the velocity vector, $t$ is time, $p$ is the pressure, and $\nu$ is the kinematic viscosity. The vector form of the momentum equation in two dimensions in the $x$ and $y$ directions is given by:

\begin{equation}\label{momentum_2D}
\begin{split}
    & u_t + uu_x + vu_y = -\frac{1}{\rho} p_x + \nu (u_{xx} + u_{yy}), \\
    & v_t + uv_x + vv_y = -\frac{1}{\rho} p_y + \nu (v_{xx} + v_{yy}).
\end{split}
\end{equation}

We define the left-hand side of the momentum equations as $\boldsymbol{f^{LHS}} = (f^{LHS}_1, f^{LHS}_2)$ where:

\begin{equation}\label{momentum_LHS}
\begin{split}
    & f^{LHS}_1 := u_t + uu_x + vu_y , \\
    & f^{LHS}_2 := v_t + uv_x + vv_y .
\end{split}
\end{equation}

The right-hand side of the momentum equations in 2D can be defined as $\boldsymbol{f^{RHS}} = (f^{RHS}_1, f^{RHS}_2)$ where:

\begin{equation}\label{momentum_RHS_2D}
\begin{split}
    & f^{RHS}_1 := -\frac{1}{\rho} p_x + \nu (u_{xx} + u_{yy}), \\
    & f^{RHS}_2 := -\frac{1}{\rho} p_y + \nu (v_{xx} + v_{yy}).
\end{split}
\end{equation}
For the 2D external and internal flows, the velocity field is reconstructed by solving the momentum equations and minimizing the residuals. Additionally, the pressure field is learned by minimizing the residual of the Navier-Stokes equations for these cases. However, in the 3D internal flow case, we reconstruct only the velocity field and do not solve for the pressure field.

\begin{figure}
    \centering
    \includegraphics[width=\linewidth]{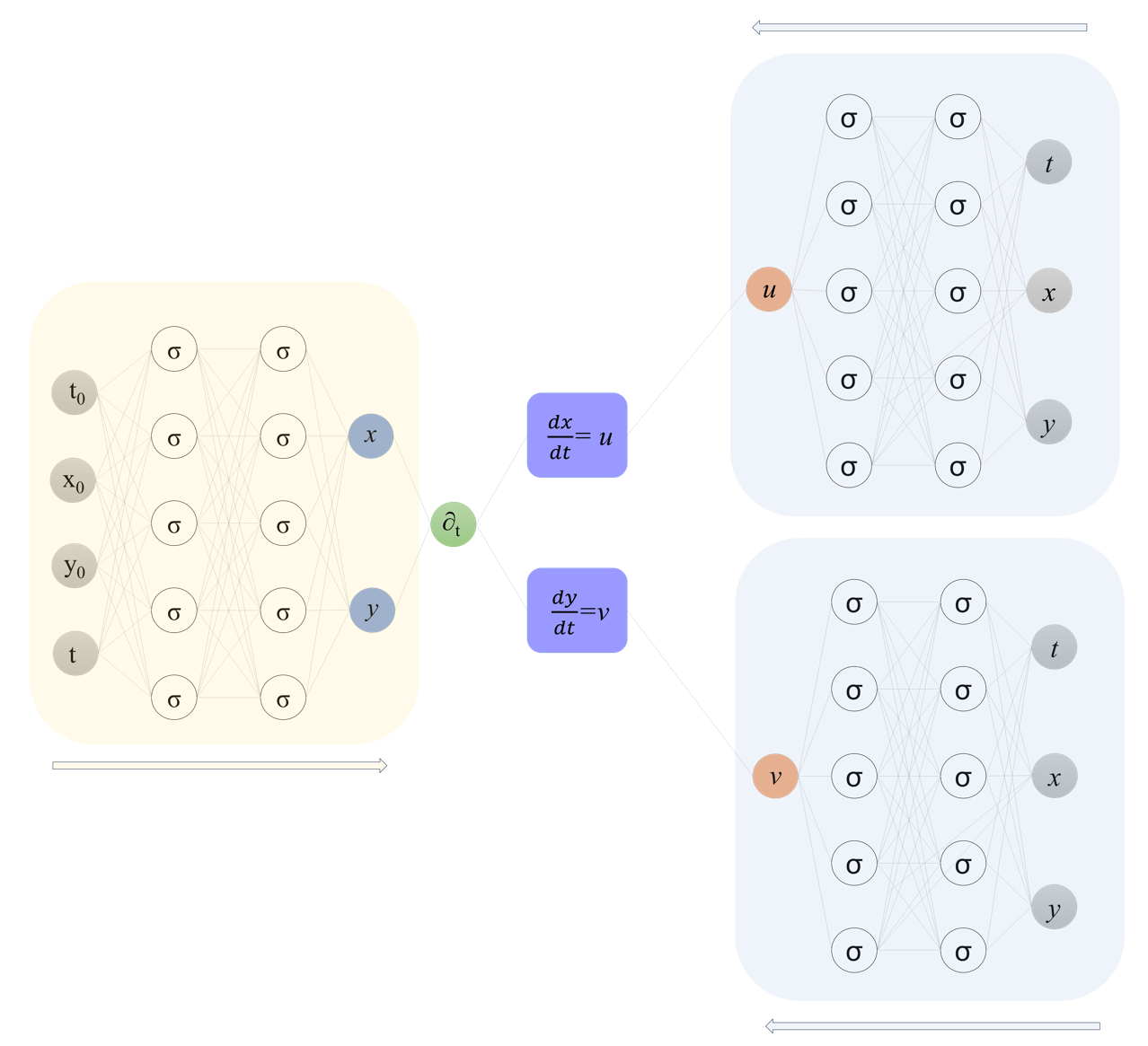}
    \caption{Neural network setup to learn the Eulerian velocity field by tracking particles in the Lagrangian domain. The arrows represent the flow of information. The peach network has the particle positions $x$ and $y$ as the ouput as a function of initial spatio-temporal location of the particle and the time as inputs. The Eulerian velocity fields $u$ and $v$ are represented as a function of $t, x$ and $y$ using fully connected networks.}
    \label{fig:fig_1}
\end{figure}
\subsection{Physics-Informed Neural Networks}

In this work, we consider the position of the particles as our only observables. To achieve this, we approximate the function $(t^0, \boldsymbol{x}^0, t) \longmapsto \boldsymbol{x}$ using a deep neural network with parameters $\theta$. Here, the superscript $0$ denotes the initial time and position of a particle, which serve as unique identifiers. Thus, the current location of a particle is modeled as a function of its initial position and time, and the current time.

We define the following mean squared loss for the regression over the particles' spatial coordinates:
\begin{equation}\label{part_data}    
    L_{\text{data}}(\theta) = \mathbb{E}_{(t^0,\boldsymbol{x}^0,t)}\left[\frac{\lvert\boldsymbol{x}(t^0,\boldsymbol{x}^0,t;\theta) - \boldsymbol{x}\rvert^2}{\sigma_{\boldsymbol{x}}^2}\right],
\end{equation}

\begin{figure}[!h]
    \centering
    \includegraphics[width=\linewidth]{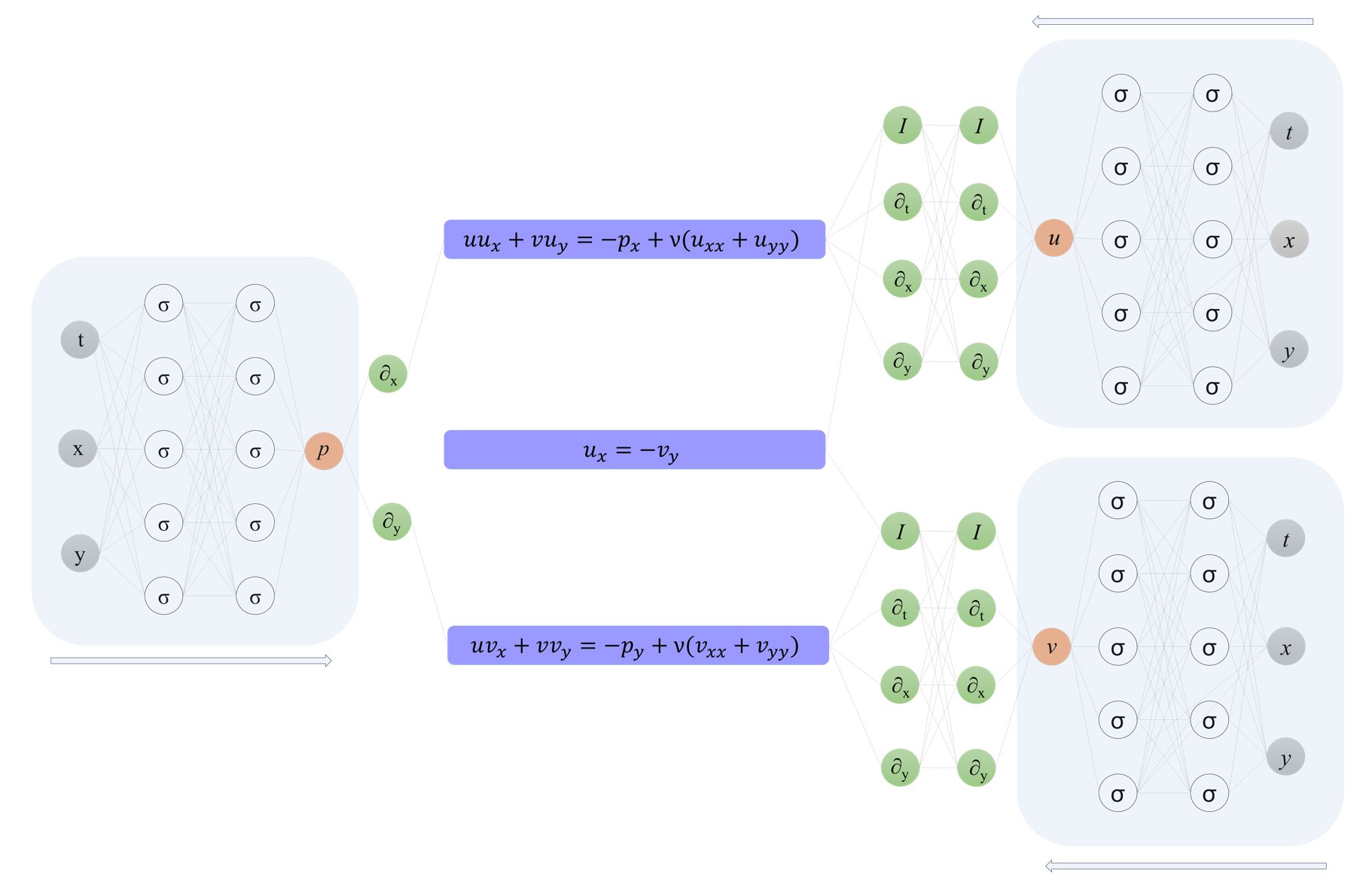}
    \caption{Neural network setup to learn the pressure field from the learned velocity fields in the Eulerian domain. The arrows represent the flow of information.}
    \label{fig:fig_2}
\end{figure}

where $\sigma_{\boldsymbol{x}}$ is the standard deviation of the reference particle position data, and $\mathbb{E}$ denotes the expectation, approximated by the sample mean. We optimize the neural network parameters $\theta^*$ by minimizing $L_{\text{data}}$, and these parameters remain fixed thereafter. This trained model allows us to compute the velocity field by differentiating the predicted position with respect to time, i.e., $\boldsymbol{x}_t$.

Next, we use two additional neural networks with parameters $\phi$ and $\kappa$ to approximate the functions $(t, \boldsymbol{x}) \longmapsto \boldsymbol{u}$ and $(t, \boldsymbol{x}) \longmapsto p$, respectively. We define the following loss function to regress the velocity field using the time derivative of the particle positions:
\begin{equation}\label{L_vel} 
    L_{\text{velocity}}(\phi) = \mathbb{E}_{(t,\boldsymbol{x})}\left[\frac{\lvert\boldsymbol{u}(t,\boldsymbol{x};\phi) - \boldsymbol{x}_t(t;\theta^*)\rvert^2}{\sigma_{\boldsymbol{u}}^2}\right],
\end{equation}

\begin{figure}[!h]
    \centering
    \includegraphics[width=\linewidth]{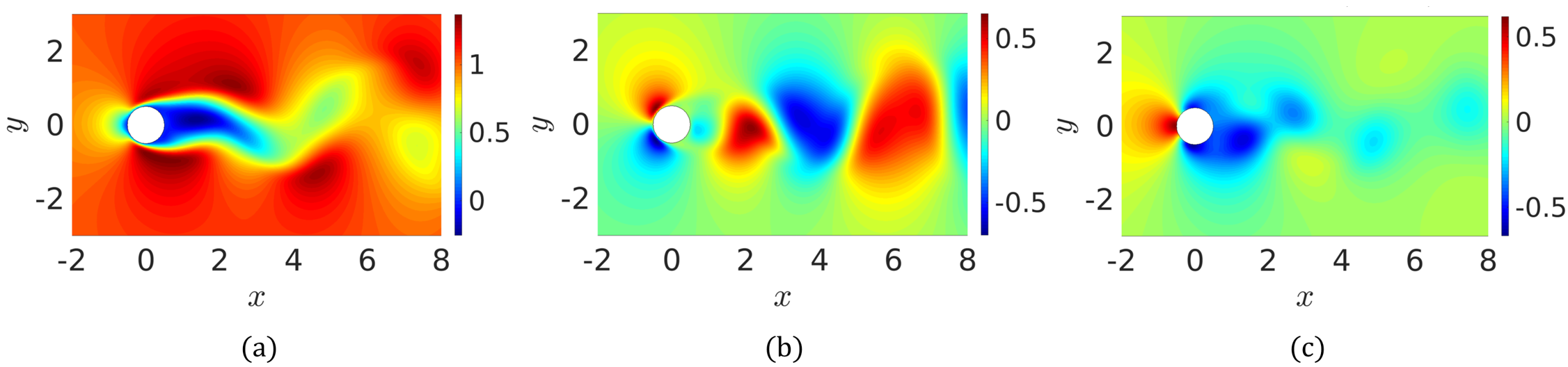}
    \caption{A snapshot of the reference (a) x-velocity, (b) y-velocity, and (c) pressure fields for flow past a cylinder.}
    \label{fig:fpc}
\end{figure}

where $\sigma_{\boldsymbol{u}}$ is the standard deviation of the velocity field estimated from $\boldsymbol{x}_t$. We optimize $\phi^*$ by minimizing $L_{\text{velocity}}$ and keep these parameters fixed in the next stage.

Finally, we infer the pressure field by minimizing the residual of the Navier-Stokes equations. We define the loss as:
\begin{equation}\label{L_NS}    
    L_{\text{N-S}}(\kappa) = \mathbb{E}_{(t,\boldsymbol{x})}\left[\frac{\lvert\boldsymbol{f}^{\text{LHS}}(t,\boldsymbol{x};\phi^*) - \boldsymbol{f}^{\text{RHS}}(t,\boldsymbol{x};\kappa)\rvert^2}{\sigma_{\boldsymbol{f}^{\text{LHS}}}^2}\right],
\end{equation}
where $\sigma_{\boldsymbol{f}^{\text{LHS}}}$ is the standard deviation of the left-hand side of the momentum equations~\eqref{momentum_LHS}.

\section{Results}\label{results}

\subsection{Case 1: Flow past a cylinder}\label{results_fpc}
To evaluate our framework and conduct a sensitivity analysis, we considered the canonical problem of flow past a cylinder, a well-known example of external flow. The reference solution was generated using \textit{OpenFOAM}~\cite{Weller198}, an open-source computational fluid dynamics toolbox. Figure~\ref{fig:fpc} shows a representative snapshot of the velocity and pressure fields from the simulation.

Inert, passive particles were generated using explicit Runge-Kutta time integration. Initial conditions for these particles were sampled by selecting random spatio-temporal points from the simulation mesh across all timesteps. A regression task was then performed on these particles by minimizing the data loss $L_{\text{data}}$ defined in Equation~\eqref{part_data}, after which the parameters of the trajectory-predicting neural network were frozen.

Using the tracked particle trajectories, the velocity field was inferred by minimizing the velocity loss $L_{\text{velocity}}$ as defined in Equation~\eqref{L_vel}. The neural network architecture used for this stage is illustrated in Figure~\ref{fig:fig_1}. Subsequently, the learned velocity field was employed to reconstruct the pressure field by minimizing the Navier-Stokes residual loss $L_{\text{N-S}}$ from Equation~\eqref{L_NS}. The neural network setup for pressure prediction is shown in Figure~\ref{fig:fig_2}.

To investigate the sensitivity of our approach to the temporal resolution of particle data, we analyzed the effect of timestep size on the reconstruction errors in the velocity and pressure fields. Table~\ref{ta:table_1} reports the corresponding errors, showing that smaller timestep sizes yield improved accuracy. The errors decreased with decreasing timestep size until $\Delta t = 0.025$~s, beyond which further reductions had negligible impact. For all cases, one million particles were used.

We further examined the sensitivity of the method to the number of particles while fixing the timestep size at 0.01~s. Doubling the number of particles from one million to two million led to negligible changes in the errors. Similarly, reducing the number of particles to 0.5 million and then to 0.2 million did not significantly impact the results. These findings suggest that our method is relatively robust to the number of particles, with reconstruction accuracy being more sensitive to the temporal resolution of the data.

{\renewcommand{\arraystretch}{1.5}
\begin{table}[ht]
  \caption{The effect of time step size on accuracy of the predictions with 1 million particles. }
  \begin{center}
    \begin{tabular}{|c|c|c|c|c|c|}
      \hline
      \bf $\Delta t$& \bf 0.01 s& \bf 0.025 s &\bf 0.05 s&\bf 0.1 s &\bf 0.2 s\\
      \hline
      $u$ & $1.87\times10^{-02}$& 2.09 $\times10^{-02}$ & $2.84\times10^{-02}$& $5.08\times10^{-02}$ & $8.91\times10^{-02}$ \\
      \hline
      $v$ & $2.24\times10^{-02}$& 2.64$\times10^{-02}$ & 3.85 $\times10^{-02}$& $7.39\times10^{-02}$ & $1.36\times10^{-01}$ \\
      \hline
      $p$ & $8.51\times10^{-02}$& 8.36 $\times10^{-02}$ & 9.28 $\times10^{-02}$& $1.36\times10^{-01}$ & $2.27\times10^{-01}$ \\
      \hline
    \end{tabular}
  \end{center}
  \label{ta:table_1}
\end{table}}

{\renewcommand{\arraystretch}{1.5}
\begin{table}[ht]
  \caption{The effect of number of particles on accuracy of the predictions for a timestep size of 0.01 seconds. }
  \begin{center}
    \begin{tabular}{|c|c|c|c|c|}
      \hline
      \bf $Paritlces$&  \bf $0.2 M$ &\bf $0.5M$&\bf $1M$ &\bf $2M$\\
      \hline
      $u$ &   $1.9\times10^{-02}$ & $1.91\times10^{-02}$& $1.87\times10^{-02}$ & $1.87\times10^{-02}$ \\
      \hline
      $v$ &  $2.23\times10^{-02}$ & 2.30 $\times10^{-02}$& $2.24\times10^{-02}$ & $2.25\times10^{-02}$ \\
      \hline
      $p$ & $8.28\times10^{-02}$ & 8.56 $\times10^{-02}$& $8.51\times10^{-02}$ & $8.47\times10^{-02}$ \\
      \hline
    \end{tabular}
  \end{center}
  \label{ta:table_2}
\end{table}}

\subsection{Case 2: Internal flow in two-dimension}\label{results_2D}

\begin{figure}[!h]
    \centering
    \includegraphics[width=\linewidth]{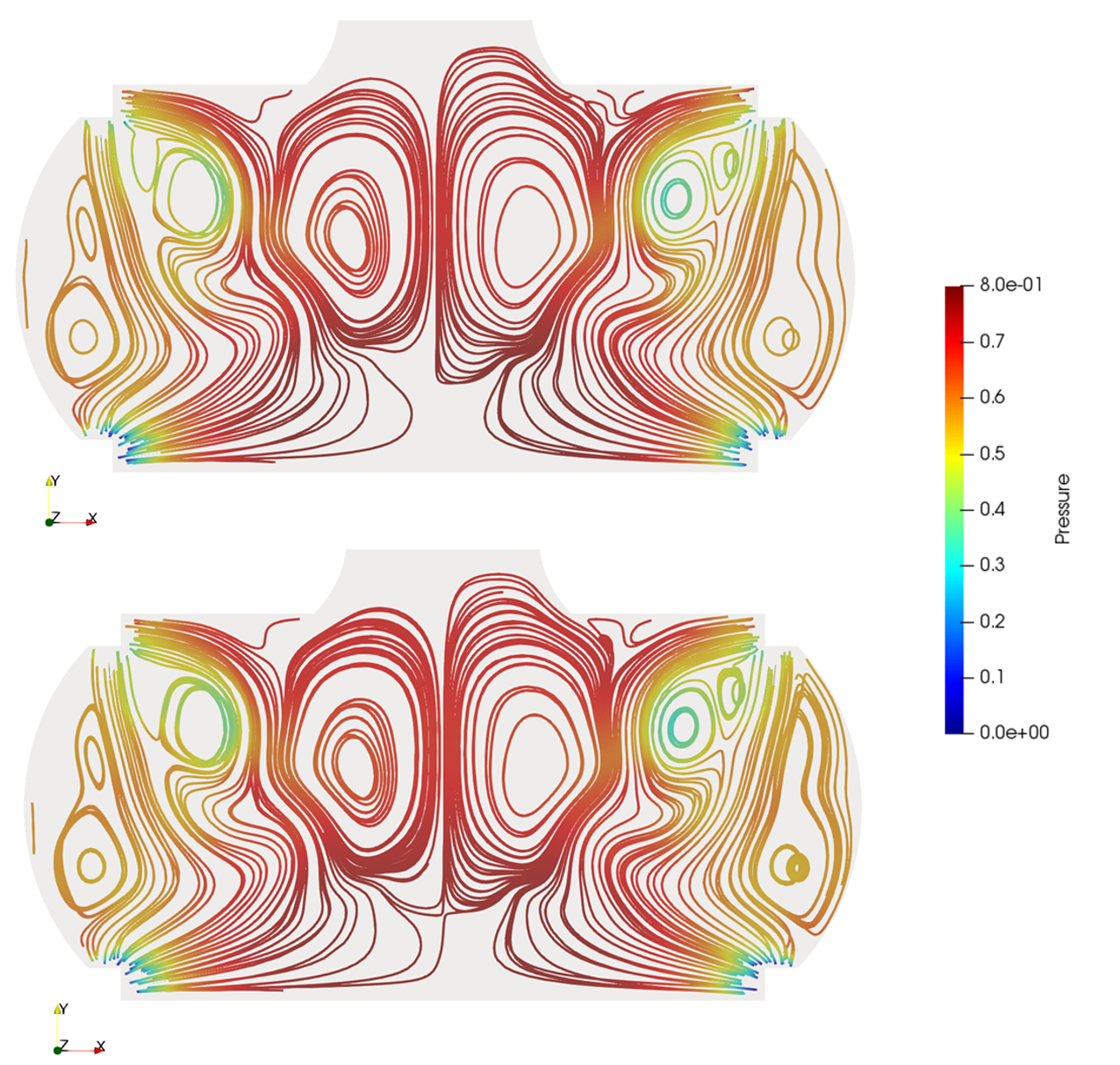}
    \caption{A comparison of the streamlines of the reference (top) and the regressed (bottom) velocity fields. The streamlines are color-coded using the reference and regressed pressure values, respectively. }
    \label{fig:airplane_2D}
\end{figure}
We now extend our framework to the task of reconstructing velocity and pressure fields in a two-dimensional internal flow scenario—specifically, airflow within an airplane cabin. A reference velocity field was generated using high-fidelity numerical simulations conducted with \textit{OpenFOAM}. This case serves as a practical and complex example of internal flow, characterized by recirculating regions, boundary interactions, and spatial non-uniformities typical of enclosed environments.

As in the external flow case, we initiated the learning process by tracking passive particles, whose trajectories were generated via explicit Runge-Kutta integration. Initial positions were sampled from the reference simulation at various timesteps, and the regression task for predicting particle positions was carried out by minimizing the data loss $L_{\text{data}}$ defined in Equation~\eqref{part_data}. Once the particle trajectories were learned, the velocity field was reconstructed by minimizing the velocity loss $L_{\text{velocity}}$ (Equation~\eqref{L_vel}) using the learned particle velocities.

The model successfully recovered the spatial components of the velocity field in both $x$- and $y$-directions with relative errors of $5.5 \times 10^{-2}$ and $2.9 \times 10^{-2}$, respectively. Using the inferred velocity field, the pressure field was subsequently reconstructed by minimizing the physics-informed loss $L_{\text{N-S}}$ in Equation~\eqref{L_NS}, achieving a relative error of $1.13 \times 10^{-1}$. Figure~\ref{fig:airplane_2D} compares the reference and predicted velocity magnitudes visualized using streamlines, color-coded by the pressure field.

The low relative errors in the velocity predictions validate the ability of our framework to effectively bridge the Lagrangian and Eulerian representations of flow. This is particularly significant in the context of internal flow applications, such as cabin ventilation, where direct velocity field measurements may be sparse or unavailable. The accurate reconstruction of airflow not only offers insights into air circulation patterns but also provides a foundation for optimizing HVAC systems and enhancing passenger comfort and safety. Furthermore, our results demonstrate that once the velocity field is accurately learned, it is possible to infer hidden quantities such as pressure purely from the underlying physics, highlighting the broader utility of our physics-informed, data-driven approach.

\subsection{Case 3: Flow in an airplane cabin}\label{results_airplane}
\begin{figure}[!h]
    \centering
    \includegraphics[width=\linewidth]{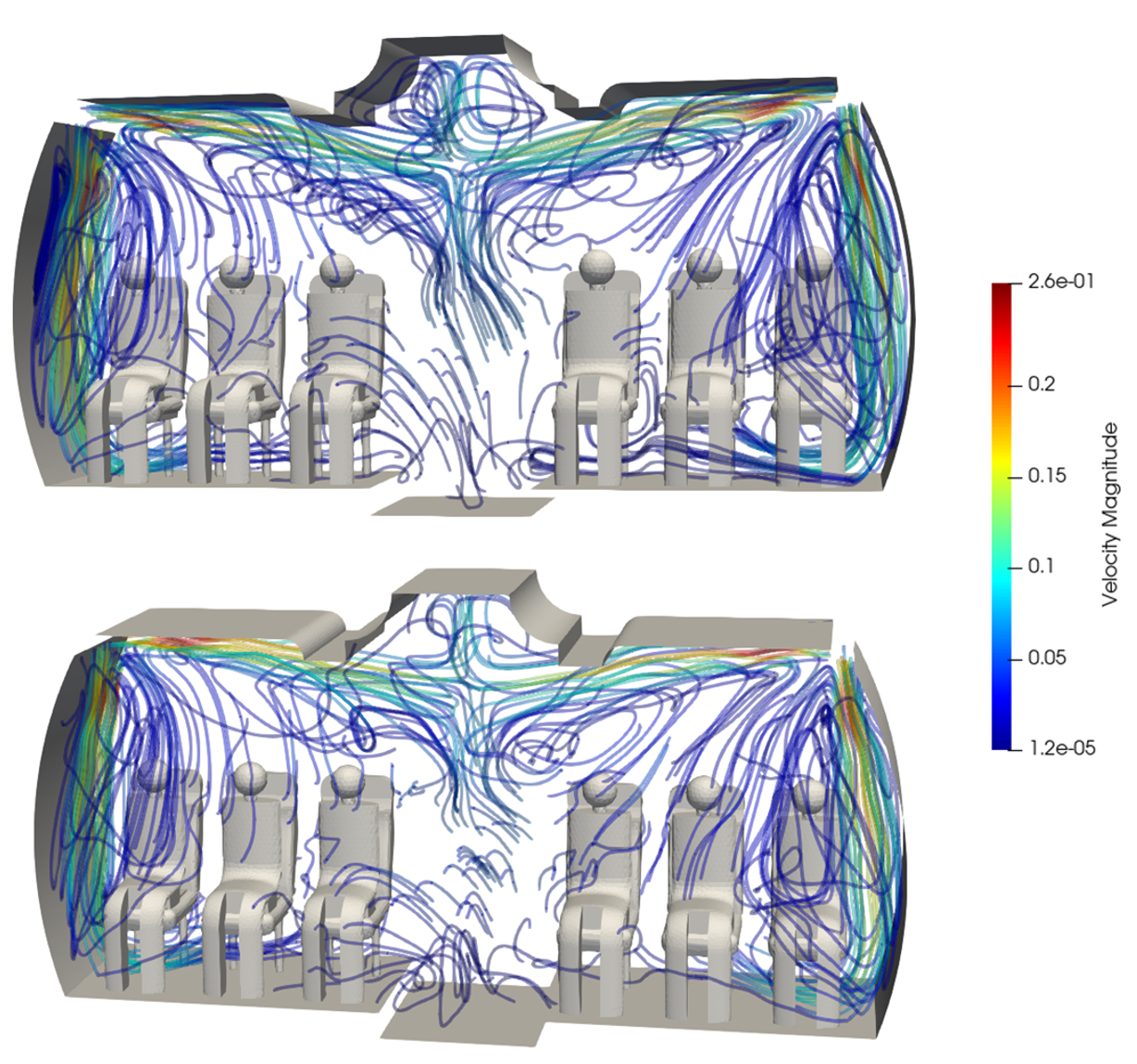}
    \caption{A comparison of the streamlines of the reference (top) and the regressed (bottom) velocity fields. The streamlines are color-coded using the velocity magnitudes. }
    \label{fig:airplane}
\end{figure}
We now apply our framework to the task of velocity field reconstruction in a complex three-dimensional internal flow scenario—specifically, airflow within an airplane cabin. A high-fidelity reference velocity field was generated through numerical simulations using \textit{OpenFOAM}. To reconstruct the flow, we tracked ten million passive particles using explicit Runge-Kutta time integration. Initial particle positions were sampled across time and space from the reference simulation to serve as inputs for the Lagrangian regression task.

Following our previous methodology, we first minimized the data loss $L_{\text{data}}$ (Equation~\eqref{part_data}) to learn the particle trajectories. The velocity field was then reconstructed by minimizing the velocity loss $L_{\text{velocity}}$ as defined in Equation~\eqref{L_vel}. Our model achieved relative errors of $5.5 \times 10^{-2}$, $2.9 \times 10^{-2}$, and $3.4 \times 10^{-2}$ in the $x$-, $y$-, and $z$-components of the velocity field, respectively.

Figure~\ref{fig:airplane} shows a comparison between the reference and predicted velocity magnitudes using streamlines, which are color-coded by the velocity magnitude. The low relative errors across all three spatial dimensions affirm the efficacy of our framework in coupling Eulerian and Lagrangian representations of flow in three dimensions.

This capability is particularly important in enclosed environments like airplane cabins, where direct measurements are sparse and understanding air circulation is critical for designing effective ventilation systems. Our results highlight that the proposed framework can robustly reconstruct velocity fields from sparse particle data in complex geometries, providing a valuable tool for scientific research and practical engineering applications. The adaptability and accuracy of our approach demonstrate its strong potential for broader use across diverse domains involving internal fluid dynamics.

\section{Conclusion and future work}\label{conclusion}

Physics-Informed Neural Networks (PINNs) have emerged as powerful tools for solving forward and inverse problems in fluid dynamics, typically within Eulerian frameworks. In this work, we propose a novel PINN-based framework that couples Eulerian and Lagrangian perspectives by leveraging particle trajectory data to infer underlying flow fields. Our method tracks the positions of inert-passive particles and reconstructs the Eulerian velocity field by solving an ordinary differential equation system consistent with the governing physics.

To validate the robustness and versatility of our framework, we tested it across three representative flow scenarios:
\begin{enumerate}
    \item \textbf{Two-dimensional external flow} past a cylinder,
    \item \textbf{Two-dimensional internal flow} in a simplified geometry,
    \item \textbf{Three-dimensional internal flow} within an airplane cabin.
\end{enumerate}

In all three cases, we successfully reconstructed the velocity field from particle tracking data. Additionally, for the two-dimensional external and internal flows, we demonstrated the ability of our framework to learn the pressure field purely from velocity information, without direct pressure measurements. The inferred pressure fields exhibited strong agreement with reference solutions, underscoring the physics-consistent nature of our approach.

Our results highlight that the framework is not only effective for reconstructing observed quantities like velocity but also capable of inferring hidden fields such as pressure through embedded physical laws. This ability makes it especially valuable for scenarios where certain measurements are difficult or impossible to obtain directly, such as in internal cavities, biomedical flows, or industrial enclosures.

Looking ahead, this framework opens promising avenues for integration with experimental particle tracking techniques and real-world applications. Potential extensions include reconstructing pressure in three-dimensional flows, inferring fluid properties like viscosity, and using sparse or noisy trajectory data. As such, our approach paves the way toward more data-efficient and physics-aware flow diagnostics, bridging the gap between simulation, experiment, and machine learning in fluid mechanics.

\bibliographystyle{elsarticle-num}
\bibliography{references}
\end{document}